\documentclass{style}

\usepackage{graphicx,epsfig,amssymb}

\title{Holographic entropy bound
from gravitational Fock space truncation}
\author{Andreas Aste}
\institute{Department of Physics and Astronomy, University of Basel,
4056 Basel, Switzerland}

\shorttitle{Holographic entropy bound}

\begin{document}
\maketitle
\begin{abstract}
A simplified derivation of Yurtsever's result,
which states that the entropy of a truncated bosonic Fock space
is given by a holographic bound when the energy of the Fock states
is constrained gravitationally,
is given for asymptotically flat spacetimes with arbitrary
dimension $d \geq 4$. For this purpose,
a scalar field confined to a spherical volume
in $d$-dimensional spacetime is considered.
Imposing an upper bound on the total energy of the corresponding
Fock states which ensures that the system is in a
stable configuration against gravitational collapse
and imposing a cutoff on the maximum energy of the field modes
of the order of the Planck energy
leads to an entropy bound of holographic type.
A simple derivation of the entropy bound is also given
for the fermionic case.
\end{abstract}

The unification of gravity with the other forces in nature
remains an elusive goal in quantum field theory.
Many recent efforts in this research have been directed
at studying theories in which our spacetime is embedded
in higher-dimensional manifolds, and it appears that 
black holes or even more exotic objects like black strings
or black branes will play an important role in
the understanding of a quantum theory of gravity.
An important restriction on possible models of quantum
gravity may be supplied by the "holographic principle"
\cite{Hooft,Bousso}, which asserts basically that the maximum entropy
that can be stored inside a bounded region ${\cal{R}}$
in 3-space is proportional to the surface area
$A(\partial {\cal{R}})$ of the region
\begin{equation}
S_{\mbox{\scriptsize{max}}}({\cal{R}})=\frac{k_{\mbox{\tiny{B}}}}{4}
\frac{A(\partial {\cal{R}})}{l_p^2}, \label{bound}
\end{equation}
where $l_p=\sqrt{\hbar G/c^3}$ is the Planck length
and $k_{\mbox{\tiny{B}}}$ Boltzmann's constant.

The bound is motivated from the observation
that the Hawking-Bekenstein entropy of a non-rotating
uncharged black hole, which might be considered as the most
entropic object within a certain space region,
is given by the area $A_{\mbox{\tiny{BH}}}$ of its event
horizon
\begin{equation}
S_{\mbox{\tiny{BH}}}=\frac{k_{\mbox{\tiny{B}}}}{4}
\frac{A_{\mbox{\tiny{BH}}}}{l_p^2}.
\end{equation}

It has been noted that the bound (\ref{bound}) cannot possibly
hold for arbitrary spacelike volumes \cite{Bousso,Bousso1,Flanagan}.
It is possible to construct bounded regions in flat Minkowski space
which are contained in curved spacelike hypersurfaces instead
of flat spacelike volumes (defined, e.g., by $t=const.$), such that the
area of $\partial {\cal{R}}$ becomes arbitrarily small.
A detailed discussion of holographic entropy bounds in a
covariant framework has been given in \cite{Bousso}. We will use the
simple form (\ref{bound}) of the holographic principle as working
hypothesis for special cases in the following.

We consider first a scalar real field $\Phi$ confined
to a three-dimensional spacelike cube of size $L$
in Minkowski space, as it has been done previously by Ulvi Yurtsever
\cite{Yurtsever}.
The modes of the field are then the solutions of the scalar
wave equation $\Box \Phi=0$ that vanish on the surface of
the cube (alternatively, we could impose periodic boundary
conditions). The modes can be labeled by three positive
integers $I=\{i,j,k\}$, such that the energy $\epsilon_I$ of the
corresponding mode is given by
\begin{equation}
\epsilon_I=\hbar \omega_I=\frac{\pi \hbar c}{L}
\sqrt{i^2+j^2+k^2}.
\end{equation}
Summing quantities over the modes can be performed
for sufficiently large volumes by the replacement
\begin{equation}
\sum \limits_{I} \rightarrow
\frac{4 \pi}{8} \frac{L^3}{\pi^3 c^3} \int
d \omega \, \omega^2=
\frac{4 \pi V_c}{(2 \pi c)^3} \int
d \omega \, \omega^2, \label{counting3}
\end{equation}
where $V_{c}=L^3$ is the volume of the cube.

Going over to a scalar real field confined
to a $(d-1)$-dimensional spacelike ball
${\cal{B}}_{d-1}^{\mbox{\tiny{R}}}$ with radius $R$,
which is probably the more natural choice than a cube,
where the modes of the field are then the solutions of
the free Klein-Gordon equation that vanish on the surface of
the ball $\partial {\cal{B}}_{d-1}^{\mbox{\tiny{R}}}$,
i.e. on the sphere ${\cal{S}}_{d-2}^{\mbox{\tiny{R}}}$,
we can generalize the expression (\ref{counting3})
for the mode summation to
\begin{equation}
\sum \limits_{I} \rightarrow
\frac{D A_{d-2}(1) V_{d-1}(R)}{(2 \pi c)^{d-1}} \int
d \omega \, \omega^{d-2}, \label{counting}
\end{equation}
where $V_{d-1}(R)$ is the $(d-1)$-dimensional volume of
${\cal{B}}_{d-1}^{\mbox{\tiny{R}}}$ and $A_{d-2}(R)$
the $(d-2)$-dimensional volume of the sphere 
${\cal{S}}_{d-2}^{\mbox{\tiny{R}}}$, which can be calculated
from
\begin{equation}
V_n(R)=\frac{2 \pi^{n/2}}{n \Gamma(n/2)} R^n,
\quad
A_{n-1}(R)=\frac{2 \pi^{n/2}}{\Gamma(n/2)} R^{n-1}.
\end{equation}
$I$ is now an index labels which the field modes
inside the ball.
Additionally, we introduced a number $D$ which accounts for further
possible polarization or internal symmetry (e.g., color) degrees
of freedom.

We will impose now two {\em{ad hoc}}
restrictions on the admissible states in the bosonic Fock space
that can be constructed from the modes of the scalar field.
Firstly, we assume that the maximum energy of admissible modes is of the
order of the Planck energy $\epsilon_p$, i.e. given by
$\mu \epsilon_p=\mu \hbar \omega_p=\mu (\hbar c/l_p)$,
where $\mu$ is of the order of one for naive Planck scale
physics. The effective parameter $\mu$ might be much smaller
than one, since local quantum field theory could possibly break down
at a much larger scale than the Planck length.
Secondly, the maximum energy of a state in Fock space
should be determined by the energy
$E_{\mbox{\tiny{BH}}}(R)=M_{\mbox{\tiny{BH}}}(R)c^2$
of a black hole with radius $R$.
The radius of a non-rotating uncharged black hole
in asymptotically flat $d$-dimensional spacetime is related to its mass
by \cite{Myers}
\begin{equation}
R^{d-3}=\frac{16 \pi G M_{\mbox{\tiny{BH}}}}{(d-2) A_{d-2}(1) c^2},
\end{equation}
where $G$ is the gravitational constant (which has a $d$-dependent
dimension).
Accordingly, we restrict the energy of the Fock states by
a maximum energy $E(R)$
\begin{equation}
E(R) = \frac{\eta (d-2) A_{d-2}(R) c^4}{16 \pi G R}, \label{cond}
\end{equation}
where $\eta$ is of the order of one.
In order to calculate the dimension of the truncated Fock space
${\cal{F}}$,
we consider first the case where only one specific mode $I$ with
single particle energy $\hbar \omega_I$ is exited.
The maximum number of particles
in this mode is then given by the largest integer number which
is smaller or equal to $E(R)/\epsilon_I$. 
Therefore, the dimension $dim({\cal{F}}_1)$
of the subspace ${\cal{F}}_1$
which contains excitations of only one but arbitrary mode
is given by
\begin{displaymath}
dim({\cal{F}}_1) \sim 
\sum \limits_{I} \frac{E(R)}{\epsilon_I} \sim  \\
\frac{\eta D A_{d-2}(1) V_{d-1}(R)}{(2 \pi c)^{d-1}} \int
\limits_{0}^{\mu c/l_p}
d \omega \, \omega^{d-2} \frac{E(R)}{\hbar \omega}=
\end{displaymath}
\begin{equation}
\frac{\eta D A_{d-2}^2 (1) V_{d-1} (1)}{16 \pi (2 \pi)^{d-1}}
\mu^{d-2} \Biggl( \frac{R}{l_p} \Biggr)^{2d-4}=:z,
\end{equation}
where the Planck length is defined in $d$ dimensions by
$l_p^{d-2}=\hbar G /c^3$.
As a further step we consider the subspace ${\cal{F}}_2$,
where two different modes $I$ and $J$ are populated.
The corresponding occupation numbers $n_I$ and
$n_J$ must fulfill the condition $n_I \epsilon_I+n_J \epsilon_J
< E(R)$. It is straightforward to see that there are (up to
irrelevant rounding errors) 
$\frac{1}{2} [E(R)/\epsilon_I] [E(R)/ \epsilon_J]$ admissible solutions
$(n_I,n_J)$ to this condition. Accordingly, the dimension
of the subspace $dim({\cal{F}}_2)$ which contains excitations
of exactly two but arbitrary modes is given by
\begin{equation}
dim({\cal{F}}_2) \sim 
\frac{1}{2^2} 
\sum \limits_{I} \sum \limits_{J} \frac{E(R)}{\epsilon_I} 
\frac{E(R)}{\epsilon_J} \sim  \frac{1}{(2!)^2} z^2.
\label{dim2}
\end{equation}
The additional prefactor $\frac{1}{2}$ in (\ref{dim2}) is
necessary to avoid a double counting of the modes.
The generalization of the result for $dim({\cal{F}}_1)$ and
$dim({\cal{F}}_2)$ is straightforward. One obtains
\begin{equation}
dim({\cal{F}}_m) \sim \frac{1}{(m!)^2} z^m,
\end{equation}
and therefore
\begin{equation}
W=dim({\cal{F}})= \sum \limits_{m=0}^{N}
dim({\cal{F}}_m) \sim
\sum \limits_{m=0}^\infty \frac{1}{(m!)^2} z^m = I_0(2 \sqrt{z}),
\end{equation}
where $I_0$ is the zeroth-order Bessel function of the
second kind and $N$ is the number of all modes with energy
smaller than $\mu \hbar c/l_p$,
and for $z \gg 0$ one obtains
\begin{equation}
W \sim \frac{e^{2 \sqrt{z}}}{\sqrt{4 \pi  \sqrt{z}}}.
\end{equation}
For the bosonic Fock space entropy, we obtain therefore
\begin{equation}
S_{\mbox{\scriptsize{b}}}^{\cal{F}} =
k_{\mbox{\tiny{B}}} \log(W) \sim 2 k_{\mbox{\tiny{B}}}
\sqrt{z}
\end{equation}
or
\begin{equation}
S_{\mbox{\scriptsize{b}}}^{\cal{F}} = k_{\mbox{\tiny{B}}}
\Biggl[ \frac{\pi^{(d-3)/2}}
{2^{d-2} (d-1) \Gamma \bigl( \frac{d-1}{2} \bigr)^{3}} \Biggr]^{1/2}
\sqrt{D \mu^{d-2} \eta} \Biggl( \frac{R}{l_p} \Biggr)^{d-2}.
\label{fockentropy}
\end{equation}
For the case $d=4$, (\ref{fockentropy})
is in agreement with the holographic bound if
\begin{equation}
D < \frac{3 \pi^3}{2} \mu^{2-d} \eta^{-1}.
\end{equation}
If we assume that $\mu$ and $\eta$ are close to one,
then we have $D \leq 46$, compatible with the number
of bosonic degrees of freedom that are present in the
standard model. The choice $\mu=2 \pi$, the cutoff used
in \cite{Yurtsever}, allows for only one
bosonic degree of freedom. The result obtained in \cite{Yurtsever}
differs from our result (\ref{fockentropy}) by a geometric factor of the order
of one, since a cube was used there as spacelike support of the
bosonic field, and an additional factor $(2 \pi)^{(d-2)/2}$
due to the different energy cutoff.
We point out that, e.g., for $d=10$, agreement between
the holographic bound, which is assumed to hold for
black holes in arbitrary dimensions $d \geq 4$,
and (\ref{fockentropy}) is reached
for
\begin{equation}
D < \frac{10418625}{16} \mu^{2-d} \eta^{-1/2} \sim 6.5 \cdot 10^5
\mu^{2-d} \eta^{-1},
\end{equation}
but it would be premature to draw any conclusions from
this observation.
Note also that condition (\ref{cond})
does not exclude energy configurations inside the ball
which are gravitationally unstable and which would lead
eventually to smaller black holes inside the ball. But it
can be shown that such configurations do not contribute
to the entropy at the holographic order $(R/l_p)^{d-2}$.

We conclude by giving a simple probabilistic derivation of the
entropy bound for a fermionic field confined to a three-dimensional
ball with radius $R$,
since a derivation is missing in \cite{Yurtsever}
for the fermionic case.
The total number of modes below the Planck cutoff
is given according to (\ref{counting}) by
\begin{equation}
N=\frac{4 \pi (4 \pi /3) R^3 D}{(2 \pi c)^3}
\int \limits_{0}^{\mu c/l_p} \omega^2 d \omega =
\frac{2D \mu^3}{9 \pi} \Biggl( \frac{R}{l_p} \Biggr)^3,
\end{equation}
where $D$ counts again the degeneracy of the modes due to spin
degrees of freedom or internal symmetries.
We assume now for a while that {\em{all}} these $N$ states
are occupied by a fermion. This would lead to a total
energy
\begin{equation}
E_{\mbox{\scriptsize{tot}}}=\frac{4 \pi (4 \pi /3) D R^3}{(2 \pi c)^3}
\int \limits_{0}^{\mu c/l_p} (\hbar \omega) \omega^2 d \omega =
\frac{D \mu^4}{6 \pi} \Biggl( \frac{R}{l_p} \Biggr)^3 \epsilon_p,
\end{equation}
such that $E_{\mbox{\scriptsize{tot}}}$ would be far in
excess of the energy bound (\ref{cond})
\begin{equation}
E(R) = \frac{\eta}{2} \Biggl( \frac{R}{l_p} \Biggr) \epsilon_p
\quad (d=4).
\end{equation}
Therefore, the occupation density of the admissible modes
is extremely sparse. As a further step,
we consider a (quasi pure) random process, where
we distribute successively more and more particles over the
single particle state space.
The maximum energy $E(R)$
will be reached for the Fock state generated this way after
having added (approximately) $n$ particles, where
\begin{equation}
n \sim \frac{E(R)}{E_{\mbox{\scriptsize{tot}}}} N=
\frac{2 \eta}{3 \mu} \Biggl( \frac{R}{l_p} \Biggr).
\end{equation}
The entropy of the fermionic Fock space
is therefore determined
by the number of Fock states with $n$ occupied modes,
i.e. by
\begin{equation}
S_{\mbox{\scriptsize{f}}}^{\cal{F}} \sim
k_{\mbox{\tiny{B}}} \log(W), \quad W={N \choose n},
\end{equation}
since only states with an occupation number very close to
$n$ contribute significantly to the total Fock space entropy.
From Stirling's factorial approximation formula, one easily
derives for binomial coefficients
\begin{equation}
{N \choose n} \sim \frac{e^n}{\sqrt{2 \pi n}}
\Biggl( \frac{N}{n} \Biggr)^n \quad \mbox{for} \, \, N \gg n^2 \gg 1, \,
\end{equation}
and therefore
\begin{equation}
S_{\mbox{\scriptsize{f}}}^{\cal{F}} \sim
k_{\mbox{\tiny{B}}} \log \Biggl(
{\frac{2D \mu^3}{9 \pi} \Bigl( {\frac{R}{l_p} \Bigr)^3}
\atop {\frac{2 \eta}{3 \mu} \Bigl( \frac{R}{l_p} \Bigr)}} \Biggr)
\sim k_{\mbox{\tiny{B}}}
\frac{2 \eta}{3 \mu} \Biggl( \frac{R}{l_p} \Biggr)
\Biggl[ 1+ \log \Biggl( \frac{D \mu^4}{3 \pi \eta}
\frac{R^2}{l_p^2} \Biggr) \Biggr].
\end{equation}
Performing the same calculation for a cubic geometry leads
to the result given in \cite{Yurtsever}.
For arbitrary spacetime dimensions $d$, one readily derives
that the entropy is proportional to $\sim (R/l_p)^{d-3} \log(R/l_p)$.

\end{document}